\documentclass[conference,10pt]{IEEEtran}

\IEEEoverridecommandlockouts

\usepackage{url}
\usepackage{cite}
\usepackage{graphicx}
\usepackage{amsmath}
\usepackage{subfigure}
\usepackage{color}
\usepackage{caption}
\usepackage{mathtools}
\usepackage{bbm}

\newcommand{\REM}[1]{}
\newcounter{todocounter}

\begin{document}

\title{End-to-End Predictions-Based Resource Management Framework for Supercomputer Jobs}

\author{\IEEEauthorblockN{Swetha Hariharan, Prakash Murali, Abhishek Pasari, Sathish Vadhiyar}
\IEEEauthorblockA{Supercomputer Education and Research Centre\\
Indian Institute of Science\\
Bangalore, India\\
swethahariharan1810@gmail.com, sercprakash@ssl.serc.iisc.in, abhishek.pasari@gmail.com, vss@serc.iisc.in}
}

\maketitle

\begin{abstract}
Job submissions of parallel applications to production supercomputer systems will have to be carefully tuned in terms of the job submission parameters to obtain minimum response times. In this work, we have developed an end-to-end resource management framework that uses predictions of queue waiting and execution times to minimize response times of user jobs submitted to supercomputer systems. Our method for predicting queue waiting times adaptively chooses a prediction method based on the cluster structure of similar jobs. Our strategy for execution time predictions dynamically learns the impact of load on execution times and uses this to predict a set of execution time ranges for the target job. We have developed two resource management techniques that employ these predictions, one that selects the number of processors for execution and the other that also dynamically changes the job submission time. Using workload simulations of large supercomputer traces, we show large-scale improvements in predictions and reductions in response times over existing techniques and baseline strategies.
\end{abstract}

\section{Introduction}

Production parallel systems in many supercomputing sites are batch systems that provide space sharing of available processors among multiple parallel applications or jobs. Well known parallel job scheduling frameworks including IBM Loadleveler\cite{IBMLL}, PBS\cite{PBS}, Platform LSF\cite{LSF} and Maui scheduler\cite{MAUI} are used to provide job queuing and execution services for users on these supercomputers. On space sharing systems, it is expected that users request a set of compute nodes for a particular duration of time. With multiple users contending for the compute resources, a batch queue submission incurs time due to waiting in the queue before the resources necessary for its execution are allocated. The queue waiting time ranges from a few seconds to even a few days on production systems. The overall response time of a user job is the sum of its queue waiting time and execution time, and is dependent on the load of the system, the batch scheduling policy and the number of processors requested by the user.

A user is presented with a unique set of challenges when using these systems for his job submission. He has to decide on the right number of processors or request size for execution. While large request sizes may result in reduction of execution times, they can cause high queue waiting times. The impact of the load conditions in the system on his job execution is not known. Even the time when he submits his job will have to be properly planned in order to obtain the best response time. In most cases, the user makes sub-optimal decisions in these aspects, resulting in high response times for the jobs and poor utilization for the system.

This paper presents an end-to-end comprehensive resource management framework that automates many of the above decision making steps. Our resource manager interfaces between the user and the underlying batch queuing system. The core components of our framework are prediction strategies for estimating the queue waiting and execution times of a job submission.
Our queue waiting time predictor uses adaptive set of strategies for choosing either distributions or summary of features to represent the system state and to compare jobs using an appropriate distance function. Depending upon the characteristics of the target and history job submissions, it varies the weights associated with the features for each job prediction, and selects a particular algorithm dynamically for performing the prediction.
Our execution time predictor is based on analyzing the impact of the load due to other executing jobs on the target job. It automatically learns this impact and forms {\em load functions}, and uses these load functions along with recent relevant history to predict a non-overlapping set of execution times, called {\em rangeSet}. The rangeSet of execution times is added with the predicted queue waiting time to form a rangeSet of predicted response times.

We have also developed two resource management strategies that use these predictions in an effective manner to minimize response times. In the first strategy, called {\em job molding}, our resource manager decides the request size for the target job by selecting from the previously used request sizes by the user such that the selected request size will most likely reduce the response time. Our second strategy called {\em delayed submissions} potentially delays the job for actual submission to the underlying queue system to a future time. In this strategy, our resource manager analyzes the impact of current and future loads on the job's execution time to make a decision.

We have evaluated our adaptive prediction framework using workload simulation traces from the Parallel Workload Archive\cite{pwa-feitelson}. Our predictions of queue waiting times result in up to 22\% reduction in the average absolute error over the existing techniques. Our predictions of execution time ranges achieve 72-89\% success rate with much lesser range lengths than the existing methods. Of our resource management strategies, the job molding strategy results in 24-53\% reduction in average response times and the delayed submission strategy yields up to 14\% additional reductions over the baseline methods.

Overall, following are the primary contributions of our work:
\begin{enumerate}
 \item We have developed an adaptive strategy for queue waiting time predictions that uses distributions to characterize the states and dynamically chooses a prediction method for each job.
 \item We have proposed a novel strategy that dynamically learns the impact of system load on the job execution time to predict a rangeSet of execution times. To our knowledge, ours is the first work that analyzes the impact of load from supercomputer traces.
 \item We have built two novel and practical resource management techniques that make decisions by comparing predicted rangeSets of response times and analyzing the impact of loads at current and future times.
 \item Finally, we have comprehensively analyzed the techniques using real supercomputing traces and queuing policies, and show large scale benefits.
\end{enumerate}

Section \ref{qwait-prediction} describes the prediction of queue waiting times, and Section \ref{exectime-prediction} describes the predictions of execution time rangeSet. In Section \ref{resource-management}, we explain our two resource management strategies of job molding and delayed submissions. In Section \ref{experimental-setup}, we describe our experimental setup including the traces used for our experiments. Section \ref{results} contains the results obtained for the predictions and the resource management techniques. In Section~\ref{related}, we compare our work with the other proposed methods on resource management in batch systems. Finally, in Section~\ref{con_fut}, we conclude and mention our plans for future work.

\section{Prediction of Queue Waiting Times}
\label{qwait-prediction}

In this section, we describe our algorithm used to predict queue waiting times in parallel systems. We present an overall view of the algorithm followed by a description of each component. A fundamental assumption in our method is that similar jobs which arrive during similar system queue and processor states experience similar queue wait times.

We use two kinds of statistics with respect to features that describe processor and queue states: feature summary and distributions. We found that using distributions help in capturing the similarities of jobs better than using feature summary if the distributions are not uniform. Using a training set, one of these two statistics is chosen. These statistics are used along with the job attributes to characterize a job at the time of its submission. We then use a weighted distance metric to calculate the similarities of the target job with the history jobs. We follow an online learning approach which uses a clustering algorithm to quickly characterize the feature neighborhood of the target job based on the distances from the history jobs. Depending on the cluster structure, we use one of three methods to calculate the predicted queue waiting time of the target job: a standard-deviation based method (SDM), nearest-neighbor method and ridge regression.

We describe the job features in Section \ref{subsection-job-features} and the distance function in Section \ref{distance-computation-subsection}. The criteria used to analyze the cluster structure of the feature neighborhood and the prediction models are presented in Section \ref{subsection-prediction-models}.

\subsection{Job Features}
\label{subsection-job-features}

At the time of arrival of a job in a supercomputer queue, certain jobs will be running on the nodes of the system and certain other jobs will be waiting in the queue. The processor state of the system contains information about the running jobs, and the queue state contains information about the waiting jobs. To predict the wait time of the new job, we look for jobs in the past which had similar resource requirements and processor and queue state as the current job. Given a job $j$, we denote the submission time of the job in the queue by $t_s(j)$, the number of nodes/cores requested by the job by $req\_size(j)$, the estimated wall clock time of the job provided by the user by $ert(j)$ (estimated run time) and the unique id of the user submitting the job by $user(j)$. These job attributes can be gathered from the job submission script provided by the user, and are also maintained in workload logs \cite{pwa-feitelson}.

We represent the system resource states using two types of statistics: distributions and feature summary. Distributions are sets of quantities associated with a particular feature. For example, the set of requests sizes of jobs waiting in the queue can be represented using a histogram distribution. Feature summary, as the name suggests summarizes the distributions to produce a representative real value. For example, a feature summary for the set of requests sizes of the waiting jobs in the queue can be the sum or average of the request sizes. To predict queue wait times for jobs at a supercomputing site, we use the job attributes in conjunction with either distributions or feature summary for the system state.

One of the important contributions of our work is the use of distributions over feature summary in some cases to represent and compare system states. In certain workloads, where distributions can reveal a distinct bias or skewness for some features, we found that the use of distributions can greatly improve the similarity computation and lead to better predictions. We use six distributions to represent the system state at the instant of arrival of a job: three distributions to represent the characteristics of the waiting jobs and three distributions to represent the characteristics of the running jobs. The distributions used by our predictor are listed in Table \ref{distribution-table}. Each distribution is represented using a histogram.

\begin{table}
  \centering
  \caption{Distributions used in distance computation}
    \begin{tabular}{|l|l|l|}
    \hline 
    No. & Type & Distribution Name                  \\ \hline \hline 
    1.  & Queue             & Request sizes of waiting jobs      \\ \hline
    2.  & Queue             & Estimated run time of waiting jobs \\ \hline
    3.  & Queue             & Elapsed wait time                  \\ \hline
    4.  & Processor         & Request sizes of running jobs      \\ \hline
    5.  & Processor         & Estimated run time of running jobs \\ \hline
    6.  & Processor         & Elapsed run time                   \\ \hline
    \end{tabular}
  \label{distribution-table}
\end{table}

When the histograms do not have  significant distinctness in the shape, we cannot rely on them to obtain a meaningful quantification of similarity. For such cases, summary of features are employed to check resource state similarity. The feature summaries and job attributes used by our predictor are listed in Table \ref{feature-table}. In the table, $\mathbbm{1}\{condition\}$ denotes the indicator function which is 1 when $condition$ is true and 0 otherwise. The first two features are the job attributes which directly influence the wait time of the job. The request size of a job is considered as a nominal attribute and the estimated run time is a numeric scalar attribute. Features 3-8 are the summarizations of the distributions listed in Table \ref{distribution-table}. One of the unique aspects of our work is that we consider user-based queue and processor features (features 9-16). These features are intended to include site specific policies which limit certain used based demands or utilization.

\begin{table*}
  \centering
  \caption{Features used in distance computation}
    \begin{tabular}{|r|l|l|} \hline
    No. & Feature name                                                             & Computation \\ \hline \hline
    1. & Request size of the current job			     &  $req\_size(J)$ \\ \hline
    2. & Estimated run time of the current job			     &  $ert(J)$ \\ \hline
    3. & Sum of request sizes of waiting jobs                        &  $\sum_{i\epsilon Q} req\_size(i)$           \\ \hline
    4. & Sum of estimated run times of waiting jobs                  &  $\sum_{i\epsilon Q} ert(i)$           \\ \hline
    5. & Sum of elapsed wait times of waiting jobs                   &  $\sum_{i\epsilon Q} (t_s(J)-t_s(i))$           \\ \hline
    6. & Sum of request sizes of running jobs                                     &  $\sum_{i\epsilon R} req\_size(i)$           \\ \hline
    7. & Sum of estimated run times of running jobs                               &  $\sum_{i\epsilon R} ert(i)$          \\ \hline
    8. & Sum of elapsed run times of running jobs                                 &  $\sum_{i\epsilon R} (t_s(J)-(t_s(i)+wait\_time(i)))$          \\ \hline
    9. & Sum of wall clock hours of waiting jobs of the current user     &  $\sum_{i\epsilon Q} \mathbbm{1}\{user(i) = user(J)\}req\_size(i)*ert(i)$          \\ \hline
    10. & Sum of request sizes of waiting jobs of the current user       &  $\sum_{i\epsilon Q} \mathbbm{1}\{user(i) = user(J)\}req\_size(i)$          \\ \hline
    11. & Sum of estimated run times of waiting jobs of the current user &  $\sum_{i\epsilon Q} \mathbbm{1}\{user(i) = user(J)\}ert(i)$          \\ \hline
    12. & Number of waiting jobs of the current user                     &  $\sum_{i\epsilon Q} \mathbbm{1}\{user(i) = user(J)\}$          \\ \hline
    13. & Sum of wall clock hours of running jobs of the current user     &  $\sum_{i\epsilon R} \mathbbm{1}\{user(i) = user(J)\}req\_size(i)*ert(i)$          \\ \hline
    14. & Sum of request sizes of running jobs of the current user       &  $\sum_{i\epsilon R} \mathbbm{1}\{user(i) = user(J)\}req\_size(i)$          \\ \hline
    15. & Sum of estimated run times of running jobs of the current user &  $\sum_{i\epsilon R} \mathbbm{1}\{user(i) = user(J)\}ert(i)$          \\ \hline
    16. &Number of running jobs of the current user                     &  $\sum_{i\epsilon R} \mathbbm{1}\{user(i) = user(J)\}$          \\ \hline
    \end{tabular}
\label{feature-table}
\end{table*}

\subsection{Distance Computation}
\label{distance-computation-subsection}

Using the features defined in the previous section, a distance function can be used to assign a real valued similarity score for a pair of jobs.
Smaller values of distance indicate higher similarity. 

\subsubsection{Distribution based Job Distance}

$\chi^{2}$ (pronounced ``chi-square'') distance metric is used to order the set of distributions of history jobs according to their similarity to the target job's distributions. 
 For two histograms $P$ and $Q$ with $K$ bins, the $\chi^{2}$ distance is defined as 
\begin{equation}
 \chi^{2}(P,Q) = \sum_{i=1}^{K}\frac{(P[i]-Q[i])^2}{P[i]+Q[i]}
\label{chi-2-distance}
\end{equation}

For each bin, the summation of bin counts in the denominator of Equation \ref{chi-2-distance} implies that $\chi^{2}$ distance considers small differences between large bins to be less important than a similar difference between small bins. Before applying the distance metric, each histogram bin is normalized by the total frequency which is same as the number of jobs involved in the histogram computation. This allows us to compare histograms of different queue and processor states although the number of jobs in each histogram may be different. Once the $\chi^{2}$ distance between a histogram of the target job and the corresponding histogram of a history job is determined for each of the six distributions shown in Table \ref{distribution-table}, the maximum of the six distances obtained is calculated. This is used to normalize the distances so that each pair of histograms has a distance in the $[0,1]$ range. The distribution based component of the distance calculation between two jobs is then defined as the sum of normalized $\chi^{2}$ distances of the six distributions.

Using the distribution distances, the final distance value for each history job is computed by adding the distances of the job attributes and applying suitable weights. Since request size is considered as a nominal attribute, $0/1$ distance is used to test whether the request sizes of the history and target jobs are the same. For estimated run time, plain difference with suitable normalization is used. This ensures that the distance value for each feature lies in a $[0, 1]$ range. For a target job $J$ and a history job $h$, the distribution based job distance is defined as:

\begin{align}
d(J,h)&= (W(feature,f_1)*\mathbbm{1}\{req(J) \neq req(j)\} \\
& + W(feature,f_2)*\dfrac{\lvert ert(J)-ert(j)\rvert}{max\_ert-min\_ert}\notag \\
& + \sum_{i=1}^{6}W(distr,d_i)*\chi^{2}(D[J][i], D[h][i]))/ W_{sum} \notag
\label{distr-job-distance-equation}
\end{align}

where W is the weighting scheme defined subsequently. $W(feature,f_i)$ and $W(distr, d_i)$ are the weights of the$i^{th}$ feature and distribution respectively. $W_{sum}$ denotes the sum of weights used for the different features. $max\_ert$ and $min\_ert$ are the maximum and minimum of the estimated run times, respectively, seen among the jobs in the history set and target job. $D[h][i]$ and $D[J][i]$ are the $i^{th}$ histograms of jobs $J$ and $h$, respectively. The overall distance value is in the $[0,1]$ range since the individual distances used in the weighted average are in the $[0,1]$ range.

\subsubsection{Feature Summary based Job Distance}

For computing the feature summary based distance between a target and history job, $0/1$ distance of request sizes and normalized plain differences of other feature values are averaged with suitable weights obtained using the weight function. For a target job $J$ and a history job $h$, the feature summary based distance is defined as:

\begin{equation}
 d(J,h) = d_n(J,h)/W_{sum}
\end{equation}

\begin{align}
d_n(J,h)&= W(feature,f_1)*\mathbbm{1}\{req(J) \neq req(h)\} \\
& + \sum_{i=2}^{16}{W(feature,f_i)*\dfrac{\lvert F[J][i]-F[h][i]\rvert}{max_{f_i}-min_{f_i}}}\notag
\end{align}

In the above equations, $F[k][i]$ denotes the value of the $i^{th}$ feature of job $k$ and $max_{f_i}$ and $min_{f_i}$ 
are the  maximum and minimum values, respectively, of the $i^{th}$ feature seen among jobs in the history set and target job.

\subsubsection{Correlation based Feature Weights}
\label{weight_function_description}

We derive weights using the correlations between features and queue waiting times for a training set of history jobs. To include the effects of dynamic policy changes by the system administrators, we recompute the weights when the history changes by addition of the latest job. We use correlation computations for calculating weights. In particular, we use the absolute values of Spearman's rank correlation coefficient, $\rho$, as weights for different features. In the case of feature summaries or job attributes like request size or estimated run time, the weight is the correlation of the feature value with the wait time of the job. For distribution based features, we choose the weight as the correlation of the $L^2$ or vector norm of the histogram
with the wait time of the job.

\subsection{Prediction Models}
\label{subsection-prediction-models}

We developed three prediction models, namely, {\em standard deviation minimizer}, {\em regression based} and {\em weighted average} methods that use the waiting times of the history jobs for predictions.  In our experiments, we found that the relative merits of the prediction models for a particular target job depend on the structure of the relationships between the waiting times and the distances in the history set. We use a density based clustering method to determine the structure of the relationships. In this section, we first describe the clustering method and then the three prediction models.

\subsubsection{Density Based Clustering}
\label{subsection-density-based-clustering}

We observed that the relationships between the waiting times of the history jobs and the distances to the target job exhibits certain characteristics which can be exploited to obtain good predictions. Specifically, we found three common patterns which allow us to choose the appropriate prediction model for a target job. Case A is where most of the jobs in the history set are far from the target job. Case B is the case where there is a dense clustering of jobs with small variance in wait time very close to the target job.  Case C is the case where there are near neighbors but they have dissimilar wait times. We can easily distinguish case A from the other two cases by checking the average distance of the closest $k\%$ of the jobs. If 
the average distance is greater than a threshold, we infer that the neighbors are too far away. In order to distinguish case B and C, we use density based clustering. 

Density based approaches view clusters as dense collections of points separated by sparse regions of low density. We use DBSCAN (Density Based Spatial Clustering of Applications with Noise)\cite{Ester96adensity-based}, a density based clustering algorithm which uses a linear number of range queries to grow clusters which maximize a density connectedness criterion. Given the wait times of the history jobs and their distance values, the cluster structure of the closest $k\%$ of the history jobs is examined. DBSCAN outputs a set of clusters along with a set of outliers. If the number of outliers is less than a threshold, our predictor assumes that there is a good clustering structure.

\subsubsection{SDM: Standard Deviation Minimizer}
\label{section-sdm}

SDM assumes that a close dense cluster of jobs exists in the neighborhood of the target job. If the cluster of jobs is dense, it implies that the jobs have comparable distance to the target job and they experienced similar queue wait times. We divide the graph of distances vs wait times for history jobs into a set of distance based windows (along the x-axis) and compute the standard deviation of wait times for jobs within each window. To obtain a very accurate prediction, a window of jobs which has minimum standard deviation and minimum distance to the target job has to be selected. Among the windows within the maximum distance threshold, the one with the smallest standard deviation is selected and the average wait time of its jobs is reported as the wait time of the target job. For computing the average of the wait time of the jobs in the cluster, we use a weighted scheme where the Gaussian kernel ($e^{d^{-2}}$ where $d$ is the distance) is used to assign higher weights for jobs with smaller distance from the target job. We pick suitable values for the various SDM parameters by performing sensitivity studies using a training set.

\subsubsection{Regression}
\label{subsubsection-regression}

When SDM cannot be applied because of poor clustering, the feature summaries of the jobs are used to construct a linear regression model. We use ridge regression since it employs a quadratic regularization term to shrink the values of the regression coefficients, making them more stable and robust to collinearity. The feature vectors are normalized to have mean 0 and variance 1 before the regression is computed. To compute the wait time for a job, the model is evaluated using target job's features. 

\subsubsection{Weighted Average}
\label{subsubsection-weighted-average}

The third method we used is k-weighted average based predictions, if the regression outputs a negative value for the wait time. To use weighted average, the distances calculated using feature summaries are used to assign weights to jobs as described in Section \ref{section-sdm}. A set of k-nearest neighbors is then chosen and the weighted average of their wait times is reported as the prediction. Based on experiments with different values of $k$, we choose a suitable value 
for each trace.

\section{Prediction of Execution Times}
\label{exectime-prediction}

\begin{figure}
\centering
\subfigure[Execution Time Variations]{
\includegraphics[scale=0.2]{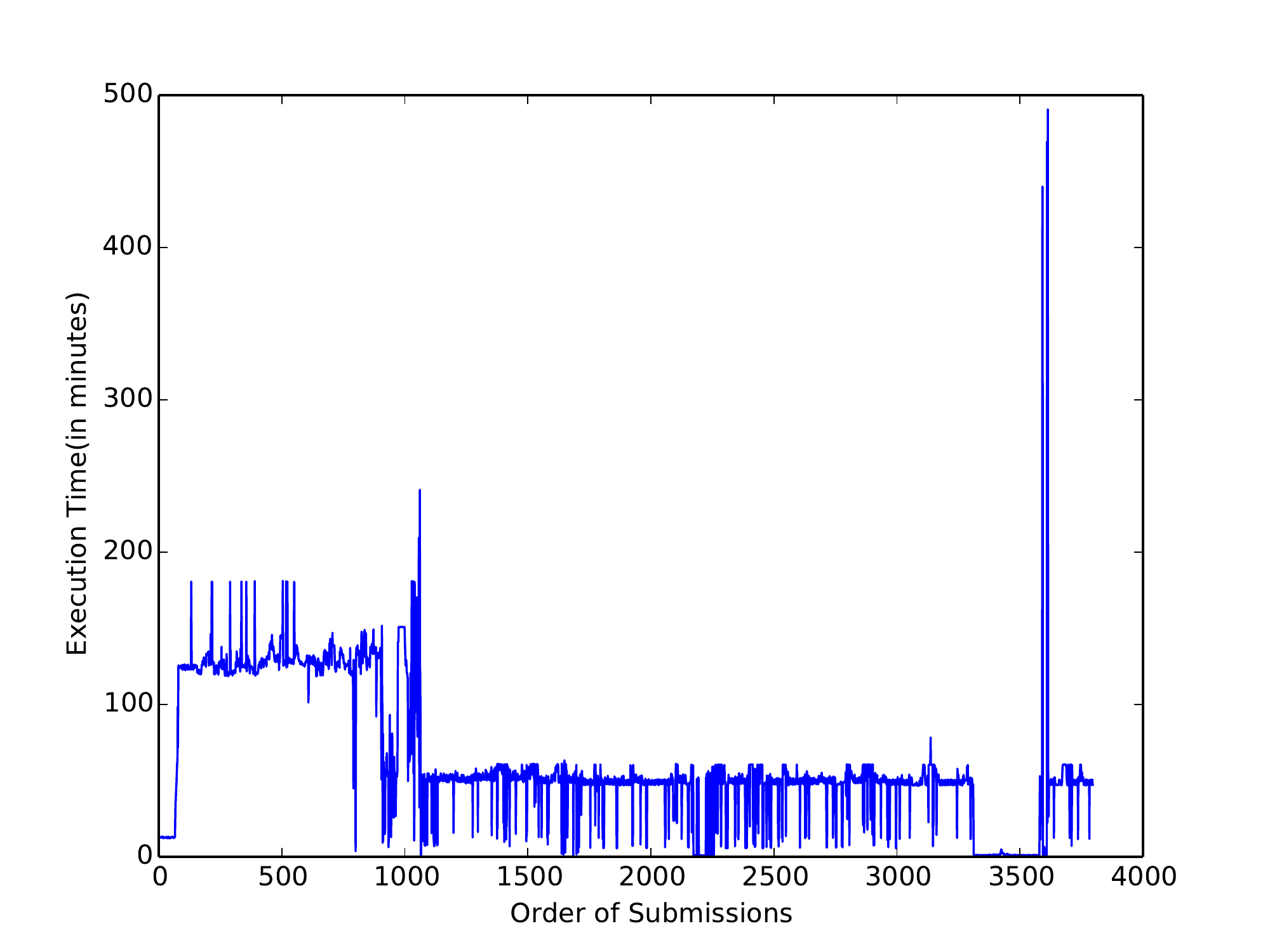}
\label{exectime-variations}
}
\subfigure[Impact of Load on Execution Times]{
\includegraphics[scale=0.2]{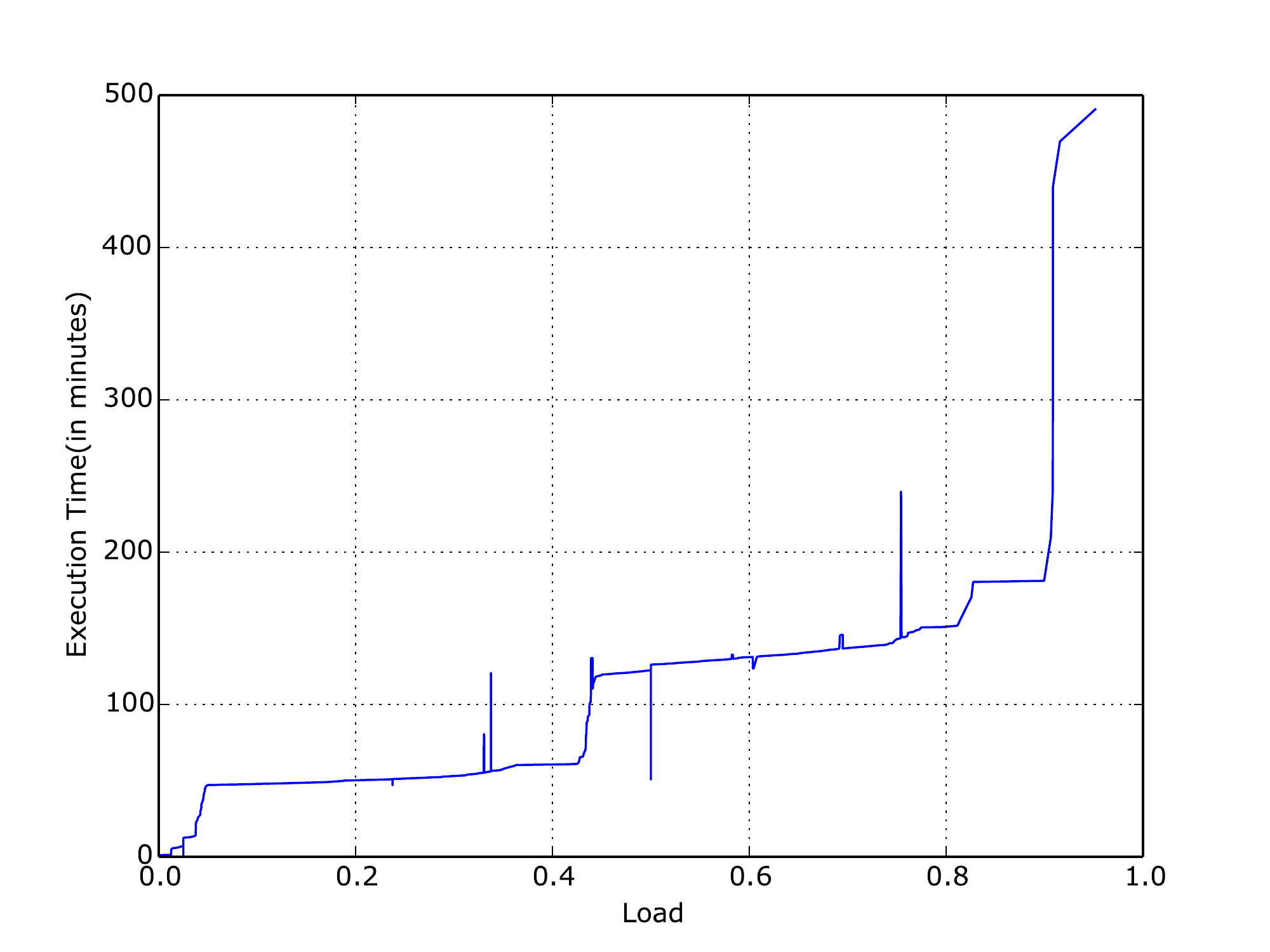}
\label{load-impact}
}
\caption{Execution Time Variations and Impact of Load for submissions with the request size of 2048 processors over a week by a user in the ANL Trace}
\label{exectime-load}
\end{figure}

One simple strategy to predict execution time of a given target job submitted by a user is to collect recent submissions by the user with the same request size and obtain the average of the execution times. This strategy assumes that all the recent submissions by the user are for the same problem and that the system exhibits similar load conditions during all the submissions. We observed in our experiments that these assumptions are not valid since the execution times for the recent submissions by the user with the same request size showed wide variations as illustrated in Figure \ref{exectime-variations}. The figure shows the execution times of a particular user with request size of 2048 processors over a one week period in the ANL (Argonne National Lab) Intrepid supercomputing trace.  While the possibility of the execution times shown in the figure belonging to different applications cannot be over-ruled, we found that the loads can have significant effect as shown in Figure \ref{load-impact}. The figure shows the execution times for the jobs arranged in increasing order of loads in the system calculated in terms of the total CPU hours utilized (Section \ref{load_prediction}). We find a good non-decreasing relationship between the execution times and the loads, thus illustrating that loads in terms of the number of other simultaneous executing jobs can have a major impact on the execution time of a job.

Hence our strategy is to predict load that may exist during the target job execution and predict execution time for the predicted load, given actual execution time and actual load for a job in the history. Since the impact of load on execution time can vary across jobs, we apply different functions of load on execution time to obtain a range of predicted execution times using a given job in the history. We consider recent job submissions, and since these submissions may belong to different applications, we obtain a set of ranges of predicted execution times. We denote this set as {\em rangeSet}.

\subsection{Calculation and Prediction of load}
\label{load_prediction}

We measure load during an interval, $I$, in terms of the amount of CPU-hours utilized by the executing applications during the interval, $T_e$, with respect to the total number of available CPU-hours during the interval, $T_t=I \times P$, where $P$ is the total number of processors in the system. The load is calculated as $T_e/T_t$.

For predicting the load during the execution of the target job, we use one-week history of workloads prior to the target job submission. We divide the one week workload into thirty minute intervals. For each interval, we obtain the loads, and use the average of the loads of the intervals as the predicted load for the interval of execution of the target job.

\subsection{Load Functions}
\label{load_functions}

To obtain the impact of load on execution times, we use a validation set and obtain a set of functions for the execution time in terms of the load corresponding to the top $k$ number of $<$$userId,requestSize,queueName,groupName$$>$ job submissions in the validation set. For each $<$$userId,requestSize,queueName,groupName$$>$ tuple, we obtain actual loads and execution times of the set of job submissions, $H$, made by the user for the request size in the validation set. If $H$ contains at least $l$ jobs and the difference between the minimum and maximum load values for jobs in $H$ is at least $m$, we consider the set $H$ as a candidate for standard function generation. We form twenty $loadBins$ with load values $[0.0-0.05], \cdots, [0.95-1.0]$. For a given load bin, we note the actual execution times of the jobs in $H$ whose actual loads fall within the load bin range. We find the average of a set of execution times in each bin and then compute the function for the average execution times in terms of the mid-points of the loads in each bin.

Since a user may have run multiple applications with the same request size with widely varying execution times, we ensure that the execution times chosen for averaging in a load bin are are within a close range. For each load bin, we use DBSCAN to find a close cluster of execution times.
While this increases the chance that the execution times chosen for averaging in a bin belong to the same application, the averages for the different bins may still correspond to different applications. We find the longest subsequence of non-decreasing averages such that the consecutive averages do not differ by more than $X\%$ thereby indicating executions of different applications. The motivation behind choosing the longest sub-sequence is to discard as few points as possible and not settle for suboptimal short sequences of runtime values. This also serves the purpose of eliminating outliers and any unrepresentative trends in the data.

\begin{figure}
\centering
\includegraphics[scale=0.35]{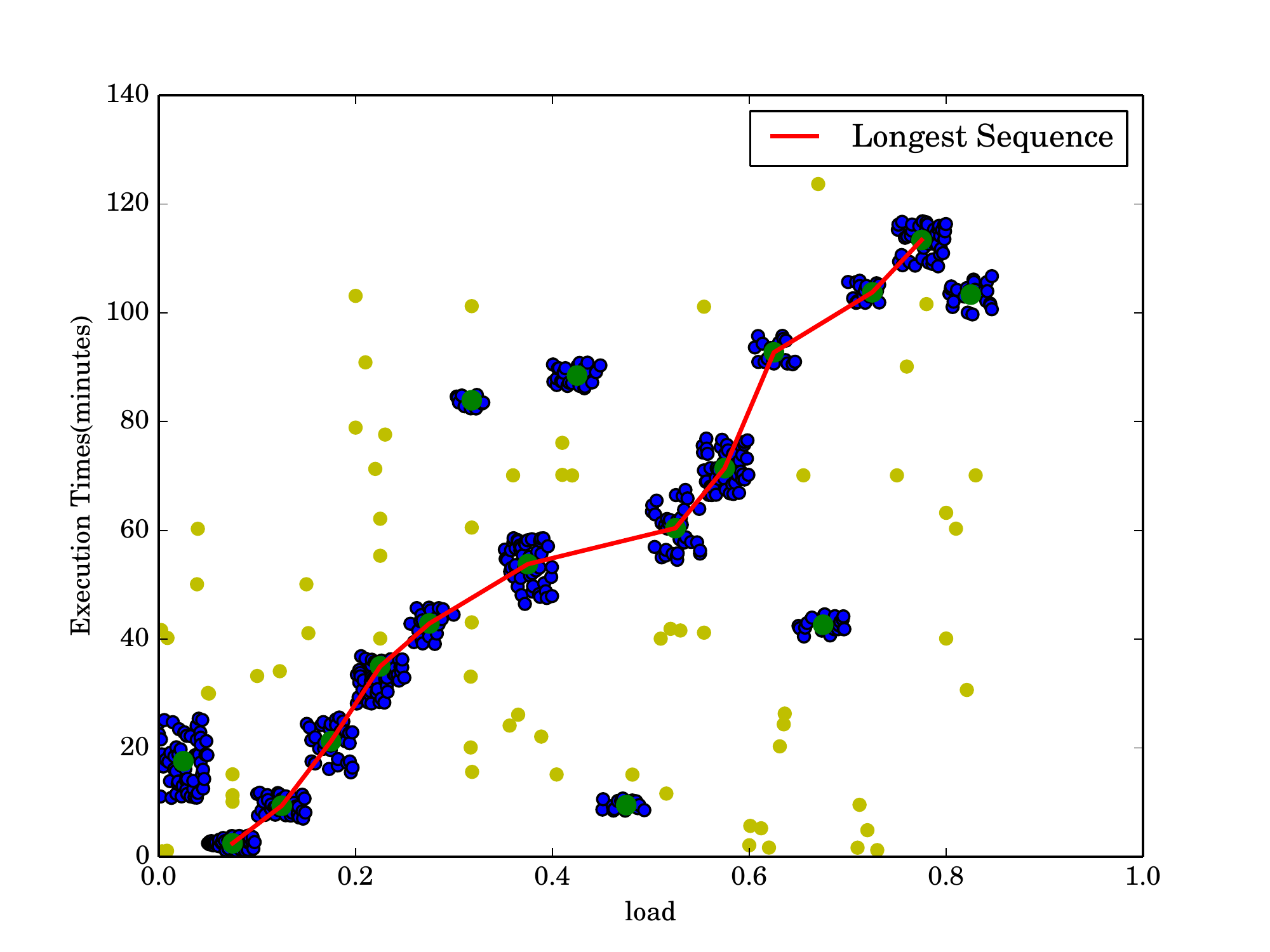}
\caption{Illustration of Finding Longest Sequence of Increasing Execution Times for Increasing Load Bins for all job submissions with the request size of 4096 processors by a single user in ANL Trace. Blue points - Clusters,  Yellow points - outliers, Green points - averages, Red line - longest sub-sequence.}
\label{longest-sequence}
\end{figure}

We formulate this problem as a longest increasing sub-sequence problem that can be solved in O($N^2$) for a total of $N$ execution times. We consider a graph based representation of the problem where each runtime value is considered as a node in a graph. A node $i$ is parameterized by the load, $l_i$, and runtime, $r_i$. Two nodes $x$ and $y$ are connected by a directed edge $(x, y)$ if for $r_x<r_y$, $l_x<l_y$ and $\frac{(r_y-r_x)\times 100}{r_x} \leq X$. The longest path in this directed acyclic graph gives the longest sub-sequence and can be solved by first linearizing the DAG and using standard dynamic programming approach \cite{dasgupta-algorithmsbook-2006} for the recurrence relation that for every node, the length of the longest path starting from the node is one more than the maximum length of longest path starting from the node's children. The process of finding the longest sequence of non-decreasing average execution times for increasing loads is illustrated in Figure \ref{longest-sequence}. The figure shows the clusters formed using DBSCAN as blue points, the outliers as yellow points, the averages in each load bin as green points, and the longest subsequence as the continuous red line.

For the selected runtime values, the load bins are now converted to point values by computing the average loads in the bins. We now find a linear function which fits these (average load, average runtime) pairs. In our implementation, we used the polyfit function in Python's numpy library \cite{numpy-web} to obtain the linear fit. The other parameters used were $k=25, l=10, m=0.5, X=50\%$.

\subsection{Relevant History for a Target Job}
\label{relevant_history}

To predict execution times for a target job submitted by a user with a request size, we consider the recent jobs submitted by the same user with the same request size. We lay emphasis on the recent job submissions by the user since the user may fine-tune his application or even change his application domain over a period of time. While choosing recent jobs, it is important to consider both the recent user job submission patterns and possible change of applications by the user in the near future. One method of obtaining recent jobs is to consider a fixed number of job submissions before the submission of the target job and whose $<$$userId,requestSize=targetJob,queueName,groupName$$>$ parameters are the same as the target job. While this considers recent jobs, it does not handle the situation where the user after submitting ten jobs belonging to an application may decide to submit a job of a different application. We propose a dynamic mechanism that strikes a balance between nearness in history and representativeness by collecting recent jobs until the execution time range of the recent set overlaps significantly (about 90\%) with the entire history set corresponding to the $<$$userId,requestSize=targetJob,queueName,groupName$$>$ tuple. We start with the recent ten jobs and traverse back in time to collect more recent jobs until this condition is met. We call this recent history as {\em relevant history} to a target job.

\subsection{Prediction of RangeSet}
\label{range_set}

We use the actual execution times and actual loads of relevant history jobs, the predicted load of the target job, and the load functions to predict a set of execution time ranges, {\em rangeSet}, for the target job. While a load function can indicate the trend of execution times with varying loads, the actual execution time for a given load can vary for different applications. For each relevant history job and a load function, we find the difference between the actual execution time of the history job and the execution time given by the load function for the actual load of the history job. We shift the function by this difference to predict the execution time for the predicted load. Application of different load functions will give a range of predicted execution times. The range that is thus obtained is compacted by adjusting the maximum and minimum limits such that the difference between them is within 20\%. The range is then added to the {\em rangeSet}.

\subsection{Overlapping to Disjoint RangeSet}
\label{merge_ranges}

The ranges thus generated in the rangeSet can contain some ranges that are overlapping. We merge these overlapping ranges such that the final rangeSet contains disjoint ranges. We have implemented a linear time algorithm for merging the overlapping ranges. We first arrange the ranges in the increasing order of minimum limits. We consider the ranges in this sequence for pushing to a stack, S. Before pushing a range in the sequence to a stack, we consider the percentage overlap between this range and the range at the top of the stack. If the percentage overlap is more than a threshold, we merge the range in the sequence with the range at the top of the stack, adjust the minimum and maximum limits of the range at the top of the stack and move to the next range in the sequence. The principle of our algorithm is that by considering the ranges in the increasing order of minimum limits, only the next range in the sequence and the top of the stack will have to be considered for merging. This is because if a range $i$ in the sorted sequence does not overlap with range $i-1$, then range $i+1$ also does not overlap with range $i-1$. The final set of disjoint ranges are contained in the stack. In our work, we fix the overlap threshold as 50\%. 

\subsection{Probability of the ranges}

We associate a probability for each range in the rangeSet by obtaining the actual execution times in the history corresponding to all job submissions whose $<$$userId,requestSize,queueName,groupName$$>$ parameters are the same as the target job, and counting the number of jobs in this history whose actual execution times belong to the range. We divide this value by the total number of job submissions to find the probability for the range. For ranges that do not correspond to any actual execution time in the history, i.e., for new ranges, we assign low probabilities such that the sum of these probabilities equals 0.1, and adjust the other probabilities for the old ranges appropriately. We also limit the total number of ranges in the rangeSet to 25. However, in practice, we obtain about 10-15 ranges.

\subsection{Putting It All Together}

\begin{figure*}
\centering
\captionsetup{justification=centering}
\includegraphics[scale=0.7]{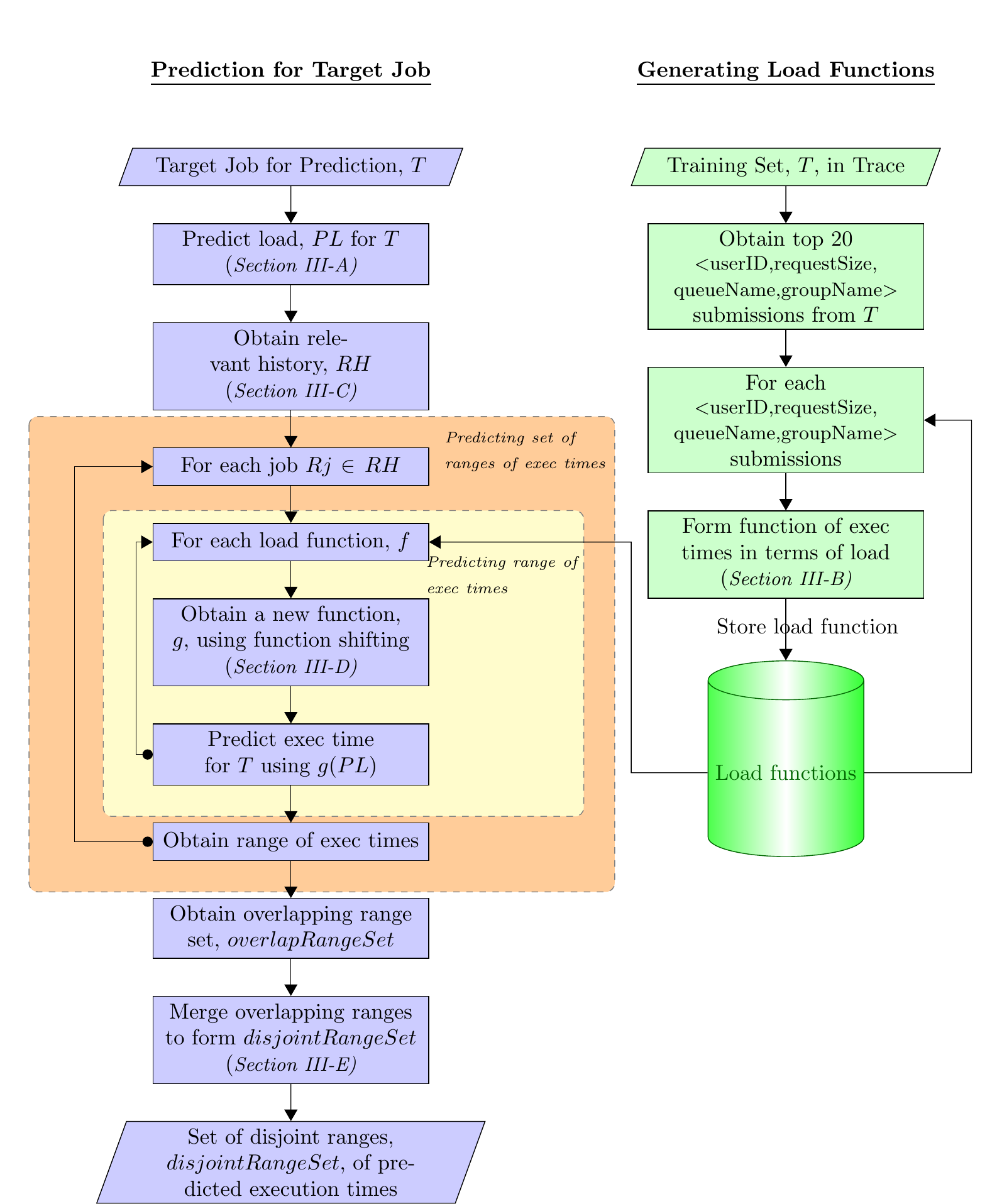}
\caption{Flowchart of Execution Time Prediction}
\label{exectime-prediction-flowchart}
\end{figure*}

The entire process of prediction of rangeSet for a target job is illustrated in Figure \ref{exectime-prediction-flowchart}.

\section{Resource Management Based on Predictions}
\label{resource-management}

It is essential to use the predictions of queue waiting, execution and response times in effective resource management strategies to optimize various metrics. We have developed two dynamic resource management techniques for minimizing response times. The first strategy is based on job molding in which the request size for a job submission is configured. The second strategy follows the process of {\em delayed submissions} in which the user job submitted to the resource management system is potentially held by the resource manager for a finite period before submitting to the actual queueing system. The following subsections describe the two strategies.

\subsection{Job Molding}
\label{job_molding}

In this scheme, for a job submitted to the system with a particular userID, the request sizes used by the user in the history of job submissions are obtained. For each such used request size, the queue waiting time and a rangeSet of execution times are predicted, as described in the previous sections. These predictions are added to obtain a rangeSet of predicted response times. The response time rangeSets corresponding to the different used request sizes are compared to select the request size that will most likely yield the least response time.

\begin{figure}
\centering
  \includegraphics[scale=0.6]{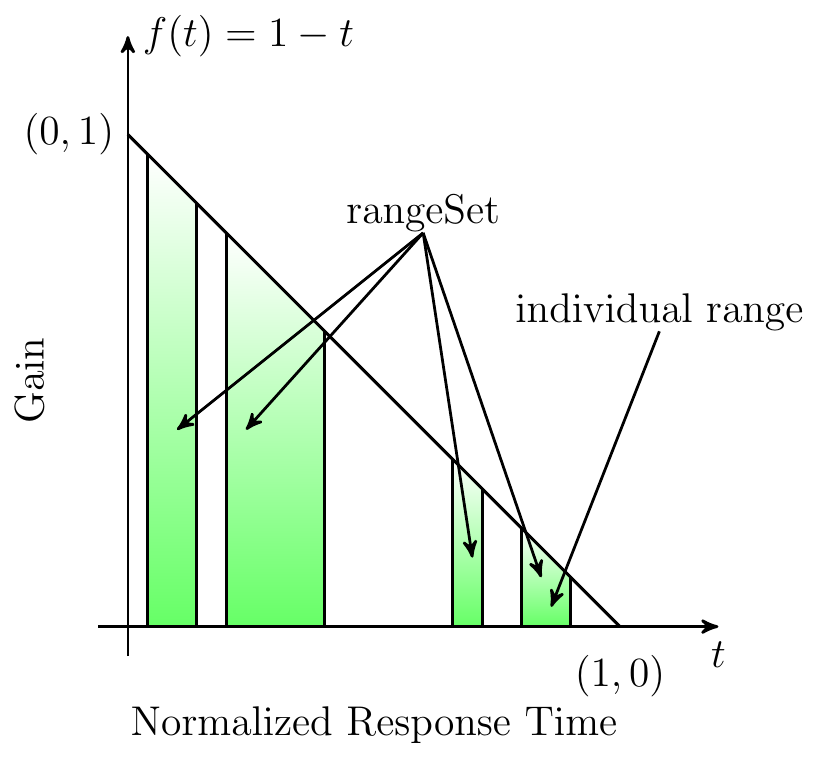}
\caption{Illustration of Gain}
\label{gain_illustration}
\end{figure}

Unlike comparing point values, comparing two different ranges or rangeSets to determine the ``smaller'' rangeSet is challenging since the ranges of the two rangeSets can overlap. We have developed a gain function for computing the score for a rangeSet. We then select the request size corresponding to the rangeSet with the highest score. We first normalize the response time ranges in the predicted rangeSet by the maximum response time so that the ranges lie between 0 and 1. We employ a cost model which assigns higher values to smaller valued predictions. Specifically, we use $f(t)=(1-t)$ as the gain function in terms of the normalized response time $t$. To compute the score or total gain for a rangeSet consisting of $N$ ranges, we compute the total area under the curve defined by the function corresponding to the $N$ ranges. This is illustrated in Figure \ref{gain_illustration}, where the shaded regions correspond to the ranges in the rangeSet. Thus, given lower and upper bounds, $l_i$ and $u_i$ for a range $i$ ($i=1\dots N$), we compute the total score as:
\begin{equation}
\label{gain_function}
score = \sum_{i=1}^N(\int_{l_i}^{u_i} f(t)\,dt.)
\end{equation}

A rangeSet which has more ranges in the higher gain region will obtain a higher score compared to a rangeSet which has more ranges in the lower gain region. To mitigate the effect of high scores obtained by accumulation of large number of small gain values corresponding to many ranges of high response times and to give higher scores for rangeSets with minimal number of ranges of low response times, we normalize the above score by the total coverage length of the rangeSet.

\subsection{Delayed Submissions}
\label{delayed_submissions}

In this scheme, in addition to reconfiguring the request size, our resource manager also reconfigures the job submission time. The jobs are first submitted to our resource manager instead of directly being submitted to the underlying batch queuing system. Our resource manager evaluates the benefit of delaying the job submission to some time in the future. This delayed submission can be useful when the system load at the current time is much higher than the load at the future time. The higher loads can result in high queue waiting times as well as in higher execution times of the job, as seen in the previous section. Hence, in some cases, the extra time delay can be very well compensated by the potential reduction in the response time of the job.

For this strategy, to determine the future time for actual job submission and to evaluate the resulting benefits, we use point predictions for execution and response times. We obtain a point prediction from a rangeSet prediction by calculating the weighted sum of the medians of the ranges in the rangeSet, where the weights equal the probabilities calculated for the corresponding ranges. For a given job submitted at a current time to our resource manager, we predict the rangeSet of predicted response times assuming the job will be actually submitted to the underlying queueing system at the current time. We denote this rangeSet as $rangeSet_{cur}$. To determine the future time, we consider the jobs that are executing in the system at the current time. Based on their runtimes predicted by our prediction strategy, we find the earliest time when one of the current executing jobs is expected to complete. Let us denote this time as $t$. Starting from $t$, we find if any other current executing job is expected to complete in a short time after $t$ ($5\%$ of $t$ in our experiments). We denote the completion time of this job as $t1$. We reset to $t$ to $t1$ and continue this search until no more job is expected to complete in a short time after $t$. We settle on this final time, $t$, as the future time. We then obtain the rangeSet of predicted response times assuming that the job will be actually submitted to the underlying queueing system at this future time. We denote this rangeSet as $rangeSet_{future}$. We obtain point predictions of response times from the two rangeSets as $predicted_{cur}$ and $predicted_{future}$. Our resource manager then delays the job for actual submission to the queueing system at the future time if $predicted_{future}$ is lesser than $predicted_{cur}$ below a threshold. In our work, we use the threshold as $30\%$. The threshold value is used to mitigate the effect of uncertainties due to predictions and score computations, future job arrivals, and to reduce the overheads of the resource management system.

To implement the delayed submissions scheme, our resource manager maintains a queue called {\em hold queue}. When the resource manager decides to delay a job submitted to it at a current time $c_t$ for actual submission to the underlying queueing system at a future time $f_t$, it holds the job in the hold queue. We denote this job as the head job. It also enqueues all the jobs that are submitted between $c_t$ and $f_t$ in the hold queue on a FCFS basis. The resource manager also notes the temporal relationships between the jobs in the hold queue by maintaining the times of submissions of these jobs to the resource manager. When the wall-clock time reaches $f_t$, the head job is actually submitted to the underlying queueing system. The rest of the jobs in the hold queue are also eventually submitted to the actual queueing system maintaining the same temporal relationships between the jobs.


\section{Experimental Setup}
\label{experimental-setup}

The experiments were conducted using real workload traces of large scale production parallel systems available from the Parallel Workloads Archive (PWA)\cite{pwa-feitelson}. Each trace, available in the Standard Workload Format(SWF)\cite{pwa-swf}, contains information about the chronology of job submissions, service times offered by the batch queue and other auxiliary characteristics which are useful for workload modeling and simulation. Job submission time along with waiting time can be used to simulate the state of the queues, and the execution start and end times of each job can be used to simulate the state of the processors. The jobs that do not execute completely due to failure or cancellation are removed from all the traces and the remaining jobs are used for our predictions. 

We have selected a set of seven traces, shown in Table \ref{supercomputing_traces} which contain sufficient information to reconstruct the queue and processor state of the system at any given time. The selected traces correspond to system sizes ranging from 128 nodes in SDSC SP2 to 163840 cores in ANL Intrepid and very low system utilization of $14.9\%$ in DAS2 to high utilization of $83.7\%$ in SDSC SP2. We also used traces of two Top500 systems - Intrepid, a BlueGene/P system with 163840 cores at Argonne National Laboratory (\#67 in Nov'13 list) and the CEA Curie supercomputer with 77184 cores (\#26 in June'14 list). Since our dataset contain widely varying system and usage profiles, we claim that our results are representative of actual production supercomputer workloads. Column 4 of the table shows the number of completed jobs we have used for our predictions.

\begin{table}
\centering
\caption{Supercomputer Traces}
    \begin{tabular}{|p{0.1in}|p{0.6in}|p{0.8in}|p{1.0in}|}
    \hline \hline
    SNo & Trace name & Trace duration (months) & Number of Completed Jobs \\ \hline \hline
    1 & ANL Intrepid & 8 & 68936 \\ \hline
    2 & CEA Curie    & 20 & 266099 \\ \hline	
    3 & DAS2         & 12 & 39915 \\ \hline
    4 & HPC2N        & 42 & 318307 \\ \hline
    5 & SDSC Blue    & 32 & 243314 \\ \hline
    6 & SDSC Paragon & 12 & 32199 \\ \hline
    7 & SDSC SP2     & 24 & 54006 \\ \hline\hline
    \end{tabular}
 \label{supercomputing_traces}
\end{table}

For workload simulation, a discrete event simulator which reads the SWF trace is used to replay the job submission, start of job execution and job termination in a chronological sequence. When a job arrives, the simulator adds the job to the list of waiting jobs in the appropriate queue. When the job begins execution on the supercomputer, the simulator moves the job from the list of queued jobs to the list of running jobs. When a job finishes running, its entry is purged from the list of running jobs. Using these lists, the simulator can maintain an online set of history job submissions which can be used for obtaining predictions for arriving jobs. The history set is updated when a currently waiting job is removed from the waiting list. When a new job entry is added to the history, the earliest entry is removed. The simulator can also interface with standard resource managers like PBS \cite{PBS}, Platform LSF \cite{LSF} or IBM Loadleveler \cite{IBMLL}. For instance, on PBS based batch queues, \textit{qstat -f} command provides the information necessary for our framework to monitor the system state.

Our predictors uses a number of parameters, e.g., history size, to tune the prediction strategy to the site and workload specific characteristics. The \textit{validation set} used for parameter tuning in queue waiting time predictions consists of 4000 jobs starting at job number 16000 in each trace. For obtaining predictions for the validation set, 6000 jobs starting from job number 10000 are considered as history submissions. We identified all the parameters used by our framework, associated each parameter with a set of possible values and varied each parameter independent of the others to find an optimal configuration of parameters on the validation set.  Table \ref{table-sensitivity-study} shows the parameters and the range of parameter values we experimented in our sensitivity studies. The optimal parameter configuration is used to obtain predictions on the \textit{test set} which consists of job numbers 20001 to 60000 in each trace. For our execution time predictions, we used a validation set consisting of the first 15000-20000 jobs to generate standard function set. The history set consists of 15,000 jobs starting from the 5000$^{th}$ job, and the test set for predictions consists of the remaining jobs starting from the 20,001$^{th}$ job. For our job molding resource management strategy, we used the first 10000 jobs as history jobs and applied job molding to the next 20,000 jobs. For our delayed submissions strategy, we used the first 10000 jobs as history jobs and applied delayed submissions to the next 6,000 jobs.

\begin{table}[h]
\centering
\caption {Parameters and Ranges of Values}
    \begin{tabular}{|p{3.5cm}|p{3.5cm}|}
    \hline
    Parameter Name                                      & Range of values \\ \hline  \hline
    Size of history set                                 & [2000-6000]     \\ \hline
    Number of bins used in the distribution             & [5-50]         \\ \hline
    Density based clustering - $k\%$, $f$, $\epsilon$, $minPts$   & [1-5], [0.10-0.90],[0.05-0.1], [2-5]  \\ \hline
    SDM - Window size, distance threshold		& [0.01-0.1],  [0.3-0.6]     \\ \hline
    Ridge regression - Maximum distance of history jobs & [0.4-1.0]       \\ \hline
    Weighted average - number of neighbors             & [1-20]          \\ \hline
    \end{tabular}
    \label{table-sensitivity-study}
\end{table}

We tested our resource management strategies using a workload and scheduler simulator that implements the EASY backfilling algorithm \cite{Feitelson:2004:PJS:2128864.2128865} to schedule jobs. We used an extended version of the Python Scheduler Simulator (PySS) developed by the Parallel Systems Lab in Hebrew University \cite{pyss}. PySS accepts a workload trace, system size and scheduling algorithm as inputs and replays the job related events to simulate the state of the system with the input workload.

\section{Results}
\label{results}

\subsection{Queue Waiting Time Predictions}

We evaluate our queue waiting time predictions on the test set using average absolute error (AAE). The average absolute error of a job is independent of the response time of the job. However, from a user's perspective an error of 20 minutes may be more acceptable for a job with response time of 10 hours than a job with response time of 100 minutes, the latter case representing a more serious prediction error. To include this bias in the error computation, we also compute the {\em scaled AAE} by dividing the AAE by the actual response time of the job. We compare the performance of our approach to two previously proposed predictors, QBETS \cite{Nurmi:2007:QQB:1254882.1254939} and IBL \cite{10.1109/CCGRID.2006.57}.

Table \ref{table-overall-prediction-accuracy} shows the AAE and the scaled AAE values for different supercomputing job traces for three methods, namely QBETS, IBL and the proposed method. We refer to our proposed method as APQ (Adaptive Prediction of Qwaits). We find that the AAE of our APQ method is up to $22\%$ and $95\%$ smaller than IBL and QBETS, respectively. We can also observe that the AAE of APQ is at least 1 hour less than IBL for SDSC SP2, HPC2N and ANL Intrepid. The scaled AAE of our APQ method is up to 375 times smaller than QBETS. In all except two cases, it is between 41\%-4.36 times smaller than IBL.
\begin{table}
\centering
\caption{Queue Waiting Time Prediction Accuracy}
\begin{tabular}{|p{0.6in}|p{0.4in}|p{0.25in}|p{0.4in}|p{0.25in}|p{0.4in}|p{0.25in}|} \hline
{Log}&\multicolumn{2}{|c|}{QBETS}&\multicolumn{2}{|c|}{IBL}&\multicolumn{2}{|c|}{APQ}\\\cline{2-7}
~&$\text{Scaled AAE}_{rt}$&AAE (hours) &$\text{Scaled AAE}_{rt}$&AAE (hours) &$\text{Scaled AAE}_{rt}$&AAE (hours)  \\ \hline \hline
ANL Intrepid	&35.15	&25.87	&0.93	&5.68	&0.55&4.52\\ \hline
CEA Curie	&2444.03&19.19	&18.21 &3.10	&20.35&2.65 \\\hline
DAS2		&32.03	&0.13		&3.13	&0.04		&1.07&0.03\\ \hline
HPC2N       	&3020.44&30.99	&40.82	&7.65	&23.29&5.97\\ \hline
SDSC Blue	&319.93	&29.44	&3.56	&5.30	&3.56&4.83\\ \hline
SDSC Paragon	&1078.39&14.99	&6.63	&0.88	&2.87	&0.69\\ \hline
SDSC SP2		&391.61	&44.37 	&51.48	&9.42	&11.81&7.84\\ \hline
\end{tabular}
\label{table-overall-prediction-accuracy}
\end{table}

\begin{figure}
\centering
  \includegraphics[scale=0.31]{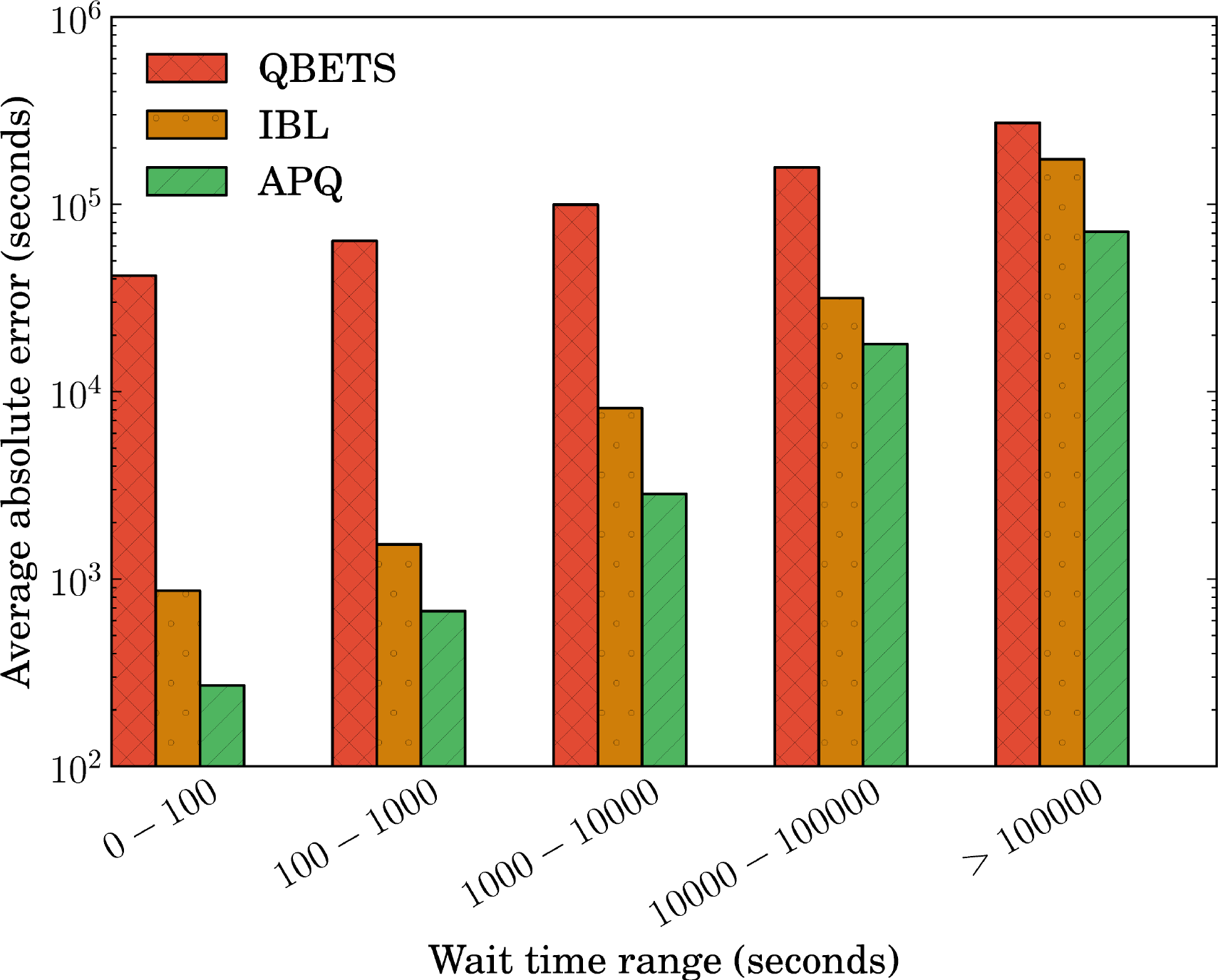}
  \caption{Prediction errors in different wait time categories}
\label{figure-prediction-error-categories-anl}
\end{figure}

In addition to showing aggregate results over all the jobs in the test set, we study the errors incurred in different job classes based on the wait time. We divide the jobs into 5 classes: $[0-100s]$, $[100-1000s]$, $...$, $[> 100000s]$ and analyze the AAE per class. Figure \ref{figure-prediction-error-categories-anl} shows the improvement in AAE obtained over IBL and QBETS for different wait time categories for ANL Intrepid trace. Since the graphs show error in log scale, small differences among the methods as seen in the graphs are significant. For example, in the $[> 100000s]$ class of Figure \ref{figure-prediction-error-categories-anl}, the AAE of our APQ method is about 4 times lesser than QBETS and 2 times lesser than IBL. Thus our method not only gives improved prediction accuracies for the overall aggregated results, but also gives improvements for different classes of jobs. 

\subsection{Execution Time Predictions}

We compare our load-based prediction strategy with three methods, namely, a {\em baseline} method and the prediction strategies by Smith et al. \cite{smith-runtimes-jpdc2004} and Li et al. \cite{li-ert-grid2005}. The {\em baseline} method predicts a range for a job by obtaining the minimum and maximum of actual execution times of job submissions in the history with the same $<$$userId,requestSize,queueName,groupName$$>$ tuple values as in the target job, and using the minimum and maximum as the bounds of the predicted range. For the methods by Smith et al. and Li et al., we used their software when available (Li et al.) or our implementation of the method (Smith et al.). In both the cases, we verified the predicted errors, obtained by executing the implementations, with the reported values.

We evaluate the ranges or rangeSets given by the prediction strategies in terms of two metrics. For a predicted range or rangeSet to be considered valuable, it must satisfy two criteria, namely {\em correctness} and {\em compactness}. We evaluate correctness in terms of percentage of successes or {\em success rate}. We define a prediction as a success if the actual execution time lies in the predicted range or rangeSet. We evaluate compactness of a range or rangeSet in terms of the total length of the range or the set of ranges. We form the ratio of this length and the length of the range predicted by the baseline method and convert the ratio to a percentage. We denote this percentage as {\em coverage}.
The methods by Smith et al. and Li et al. give give point predictions. To convert a point prediction for a target job from one of these methods to a range prediction, we obtained the average absolute error (AAE) of the predictions with the method for all jobs in the history with the same  $<$$userId,requestSize,queueName,groupName$$>$ tuple values as the target job, and calculate the range of length equal to AAE with the actual execution time of the target job as the median of the range.

Table \ref{exectime-predictions-rangecomparisons} shows the comparisons in terms of success rate, $SR$ and coverage, $C$. Note that the coverage value of the baseline method is 100\% since the coverage is calculated using the results of the baseline method. We find that our method achieves a success rate of 72-89\% with an average of 80\% for the predictions. While the success rate with our method is 2-16\% with an average of 6.5\% lower than the success rate of the baseline method, we were able to achieve our high success rates with only 7-26\% coverage, i.e., the ranges by our method are 74-93\% more compact than the baseline ranges. We find that the success rates of the Smith et al.'s and Li et al.'s methods are smaller than our method, while their coverages are high percentages with the lengths of their predicted ranges much larger than the ranges by the baseline.

\begin{table}
\centering
\caption{Execution Time Range Prediction Accuracy}
\begin{tabular}{|p{0.59in}|p{0.31in}|p{0.15in}|p{0.15in}|p{0.15in}|p{0.15in}|p{0.15in}|p{0.15in}|} \hline
Log & Baseline & \multicolumn{2}{|p{0.45in}|}{Smith et al.} & \multicolumn{2}{|p{0.45in}|}{Li et al.} & \multicolumn{2}{|p{0.5in}|}{Our Method} \\ \hline
 & SR & SR & C & SR & C & SR & C \\ \hline\hline
ANL Intrepid	& 89 & 67 & 156 & 71 & 102 & 84 & 17 \\ \hline
CEA Curie	& 88 & 58 & 218 & 62 & 206 & 72 & 23 \\ \hline
DAS2		& 95 & 78 & 120 & 83 & 98  & 89 & 7 \\ \hline
HPC2N       	& 83 & 63 & 166 & 72 & 140 & 77 & 19 \\ \hline
SDSC Blue	& 85 & 67 & 326 & 71 & 302 & 83 & 14 \\ \hline
SDSC Paragon	& 89 & 70 & 145 & 75 & 113 & 78 & 8 \\ \hline
SDSC SP2	& 79 & 69 & 339 & 76 & 213 & 81 & 26 \\ \hline\hline
\end{tabular}
\label{exectime-predictions-rangecomparisons}
\end{table}

The point predictions by Smith et al. and Li et al. have average percentage prediction errors of at least 40\%. While we believe that providing ranges will be more useful to the user than providing point predictions with such high errors, we nevertheless obtain point predictions from our rangeSets for fair comparisons with the point predictions given by the other two methods. To obtain point predictions for our method, we obtain a weighted sum of the median values of the ranges in the set with the weights equal to the probabilities of the ranges. We note that this is a gross over approximation of the rangeSet into a point value. Table \ref{exectime-predictions-pointcomparisons} shows the comparisons of the point predictions. We find that even with the above gross approximation, our method gives point predictions with lower AAEs than the other two methods for five of the seven traces.

\begin{table}
\centering
\caption{Execution Time Point Prediction Accuracy}
\begin{tabular}{|p{0.8in}|p{0.5in}|p{0.5in}|p{0.5in}|} \hline
Log & AAE by Smith et al. (minutes) & AAE by Li et al. (minutes) & AAE by Our method (minutes) \\ \hline\hline
ANL Intrepid	& 50 & 48 & 43 \\ \hline
CEA Curie       & 66 & 76 & 116  \\ \hline
DAS2	        & 25 & 19 & 18 \\ \hline
HPC2N           & 46 & 67 & 64 \\ \hline
SDSC Blue	& 44 & 48 & 41 \\ \hline
SDSC Paragon	& 41 & 58 & 36 \\ \hline
SDSC SP2	& 56 & 69 & 43  \\ \hline\hline
\end{tabular}
\label{exectime-predictions-pointcomparisons}
\end{table}

\subsection{Resource Management: Job Molding}

We compare our job molding strategy to a baseline method. We evaluate these strategies using the EASY backfilling scheduler simulator of PySS. The scheduler simulator requires a job trace containing for each job, the job submission time, requested number of processors, expected runtime and the actual runtime. The simulator follows the scheduling policy to allocate the jobs for executions and outputs the queue waiting times of the jobs. We use our seven supercomputer job traces as inputs to the scheduler simulator. For a given job trace, we use its job submission time in both the job molding and baseline strategies. While we use the same request sizes, expected runtimes and actual runtimes as given in the trace for the baseline method, we alter these parameters for the job molding strategy. The request size in the job molding strategy is determined as described in Section \ref{job_molding} on job molding. We denote this request size and the one as given in the trace as {\em changed} and {\em original} request sizes, respectively.

For setting the estimated runtime for job molding, we obtain the estimated runtimes of all previous job submissions submitted by the same user for the changed request size and find the maximum of these times. We denote this maximum as $maxPrevious$ and the maximum runtime limit in the supercomputer system as $maxLimit$. If the changed request size is less than the original request size, we expect the estimated runtime for job molding to be greater than the original estimated runtime. In this case, we set the estimated runtime as the minimum value of $maxPrevious$ and $maxLimit$ that is greater than the original estimated runtime. If the changed request size is greater than the original request size, we expect the estimated runtime for job molding to be at most the original estimated runtime. In this case, we set the estimated runtime as the original estimated runtime. For setting the actual execution time for job molding, we obtain the actual runtimes of all previous job submissions submitted by the same user for the changed request size and probabilistically choose one of the runtimes with a Roulette wheel selection strategy.

Table \ref{jobmolding-results} shows the average queue waiting, execution and response times in minutes with the baseline and our methods. We find that our job molding strategy results in the reduction of both queue waiting and execution times, and hence reduction in response times over the baseline method. Overall, the job molding strategy results in 24-53\% reduction in average response times. Thus, our predictions can be effectively used in a simple and practical strategy like job molding to obtain large benefits.

\begin{table}
\centering
\caption{Average Queue Wait, Execution and Response Times for Job Molding Strategy (All times in Minutes)}
\begin{tabular}{|p{0.6in}|p{0.2in}|p{0.15in}|p{0.3in}|p{0.2in}|p{0.15in}|p{0.3in}|} \hline
Log & \multicolumn{3}{|p{0.65in}|}{Baseline} & \multicolumn{3}{|p{0.65in}|}{Our Method} \\ \hline
 & QWait [A] & Exec. [B] & Response [C=A+B] & QWait [A] & Exec. [B] & Response [C=A+B] \\ \hline\hline
ANL Intrepid	& 61 & 83 & 144 & 26 & 41 & 67 \\ \hline 
CEA Curie       & 15 & 95 & 110 & 8 & 64 & 72 \\ \hline 
DAS2	        & 0.12 & 6 & 6.12 & 0.02 & 4.17 & 4.19 \\ \hline 
HPC2N           & 72 & 289 & 361 & 50 & 210 & 260 \\ \hline 
SDSC Blue	& 97 & 73 & 170 & 72 & 56 & 128 \\ \hline 
SDSC Paragon	& 102 & 79 & 181 & 76 & 51 & 127 \\ \hline 
SDSC SP2	& 198 & 133 & 331 & 147 & 98 & 245 \\ \hline\hline 
\end{tabular}
\label{jobmolding-results}
\end{table}

\begin{figure}
\centering
  \includegraphics[scale=0.4]{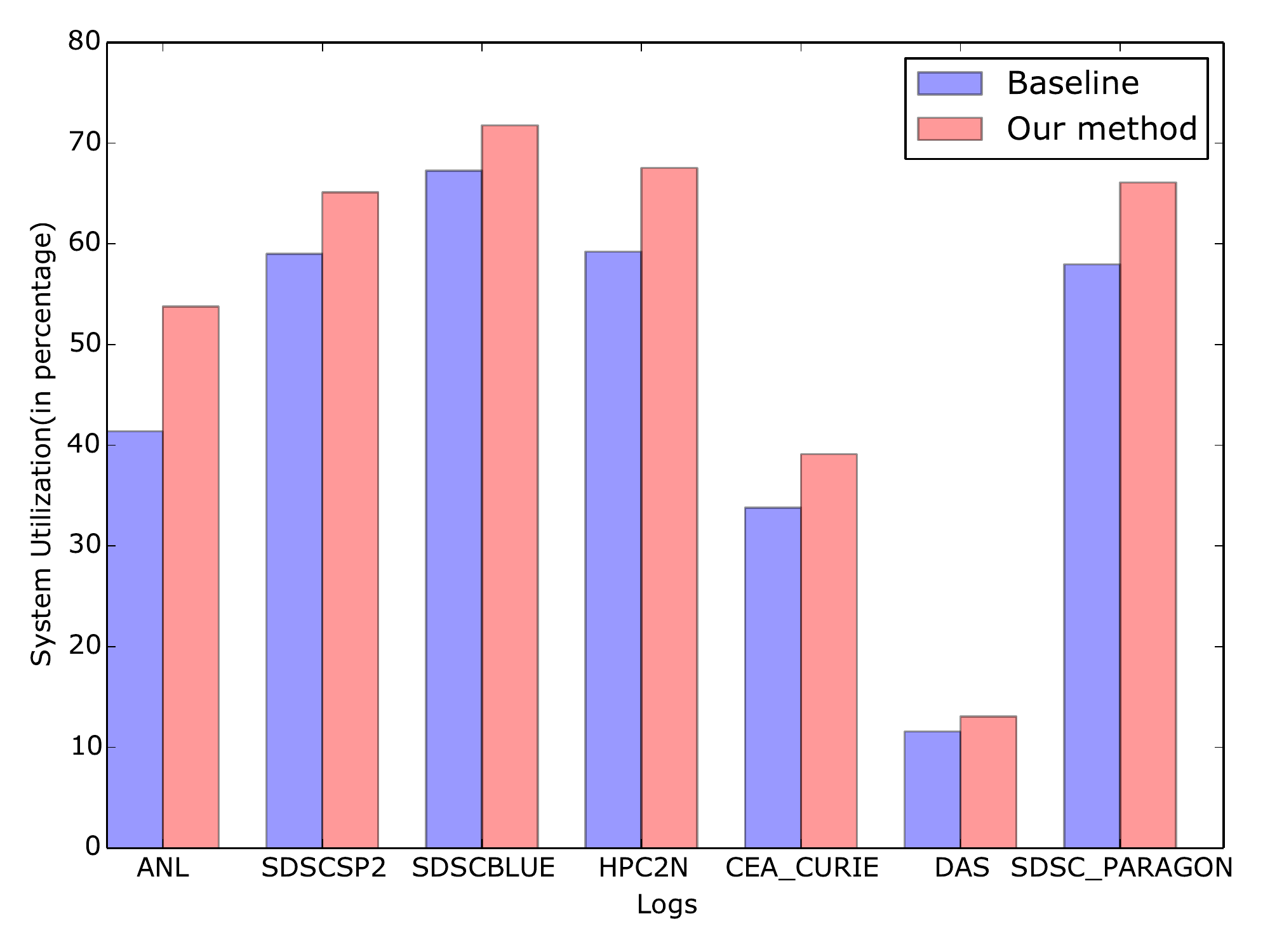}
\caption{System Utilization with the Job Molding Strategy}
\label{util_jm}
\end{figure}

Figure \ref{util_jm} shows the total system utilization, calculated as the percentage of CPU hours utilized for executions, obtained on the seven systems using the baseline method and our strategy. We find that job molding, based on predictions of queue waiting and execution times, results in 10-15\% increase in system utilization. When the system is highly loaded, the baseline method will waste the CPU hours due to insufficient free resources for executing the jobs with the fixed request size provided by the users. This also results in high queue waiting and hence high response times for the jobs. Our job molding strategy potentially reduces the request size of the job if it can result in reduced queue waiting and response times. This allows the jobs to be efficiently packed into the available resources.

\begin{figure}
\centering
  \includegraphics[scale=0.4]{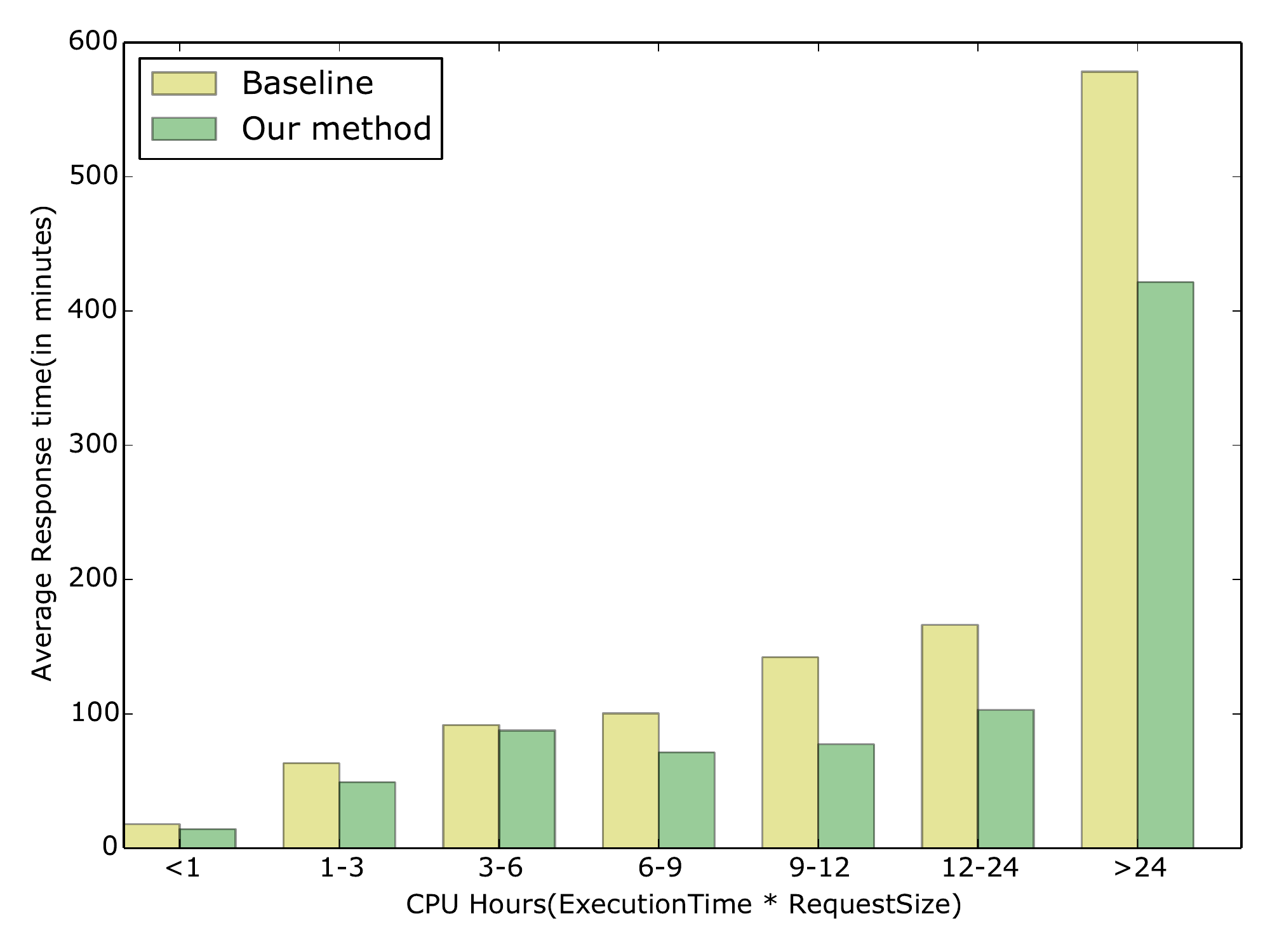}
\caption{Response Times for Jobs of Different Sizes for Job Molding in SDSC Blue}
\label{cpuhours_response_jm}
\end{figure}

Figure \ref{cpuhours_response_jm} shows the average response times with the baseline and job molding strategies for jobs of different sizes corresponding to different CPU hours in the SDSC Blue trace. We find that our job molding strategy uniformly provides benefits across all the sizes of jobs. This is due to the adaptation of our strategy to system loads, expanding the small-sized jobs in case of light loads, and shrinking the large-sized jobs in case of high loads.

\begin{figure}
\centering
  \includegraphics[scale=0.4]{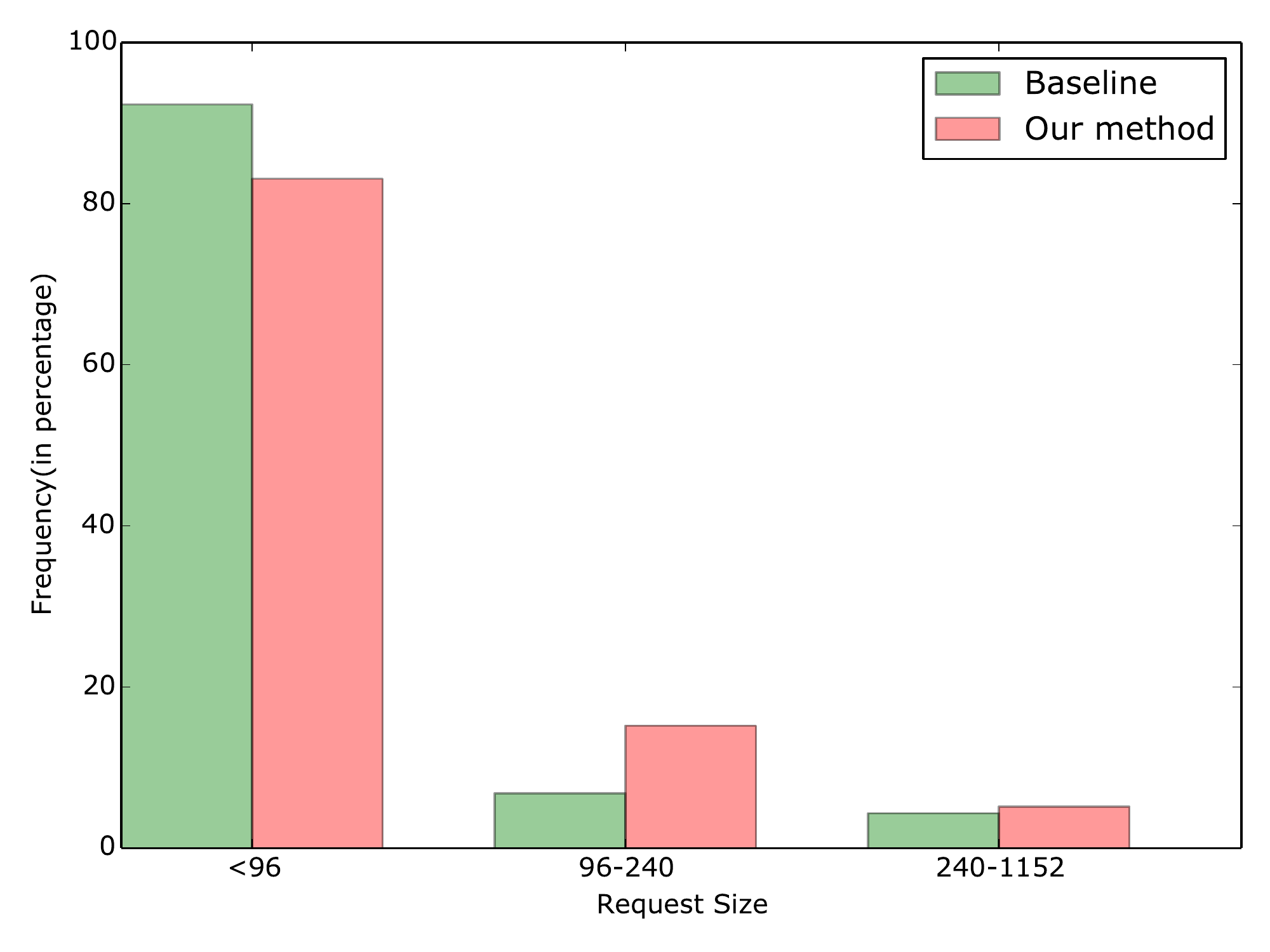}
\caption{Distribution of Request Sizes for Job Molding in SDSC Blue}
\label{rs_freq_jm}
\end{figure}

Figure \ref{rs_freq_jm} shows the overall distribution of request sizes achieved by the strategies for the SDSC Blue Trace. We find that in general, the job molding strategy tries to provide better response times by expanding many small-sized jobs to large-sized jobs, when resources are available. This results in reduction in execution times and hence the response times of these jobs.

\subsection{Resource Management: Delayed Submissions}

We compared our resource management strategy of delayed submissions to the same baseline method described in the previous section. The delayed job submission time to the EASY backfilling scheduler simulator is determined using the procedure described in Section \ref{delayed_submissions}. The request size is changed as in job molding considering the load of the system at the delayed job submission time. This load is expected to be smaller than the load at the original job submission time due to expected completion of some of the running jobs. We calculate this load as $futureLoad$. The expected and actual runtimes of the delayed job are calculated at the $futureLoad$ using the same strategy as in job molding.

Table \ref{delayedsubmission-results} shows the averages of the various times in minutes with the baseline and our methods. We find that our delayed submissions strategy results in the reduction of response times over the baseline method. Overall, the delayed submissions strategy results in 30-50\% reduction in average response times, and yields up to 14\% additional reduction to using job molding alone. 

\begin{table}
\centering
\caption{Average Queue Wait, Execution and Response Times for Delayed Submissions Strategy (All times in Minutes)}
\begin{tabular}{|p{0.6in}|p{0.2in}|p{0.15in}|p{0.3in}|p{0.2in}|p{0.15in}|p{0.3in}|} \hline
Log & \multicolumn{3}{|p{0.65in}|}{Baseline} & \multicolumn{3}{|p{0.65in}|}{Our Method} \\ \hline
 & QWait [A] & Exec. [B] & Response [C=A+B] & QWait [A] & Exec. [B] & Response [C=A+B] \\ \hline\hline
ANL Intrepid	& 43 & 76 & 119 & 26 & 57 & 83 \\ \hline 
CEA Curie       & 11 & 84 & 95 & 5 & 50 & 55 \\ \hline 
DAS2	        & 0.01 & 2.24 & 2.25 & 0 & 1.11 & 1.11 \\ \hline 
HPC2N           & 68 & 265 & 333 & 47 & 180 & 227 \\ \hline 
SDSC Blue	& 91 & 65 & 156 & 63 & 44 & 107 \\ \hline 
SDSC Paragon	& 114 & 89 & 203 & 71 & 42 & 113 \\ \hline 
SDSC SP2	& 200 & 142 & 342 & 153 & 79 & 232 \\ \hline\hline 
\end{tabular}
\label{delayedsubmission-results}
\end{table}

\begin{figure}
\centering
  \includegraphics[scale=0.4]{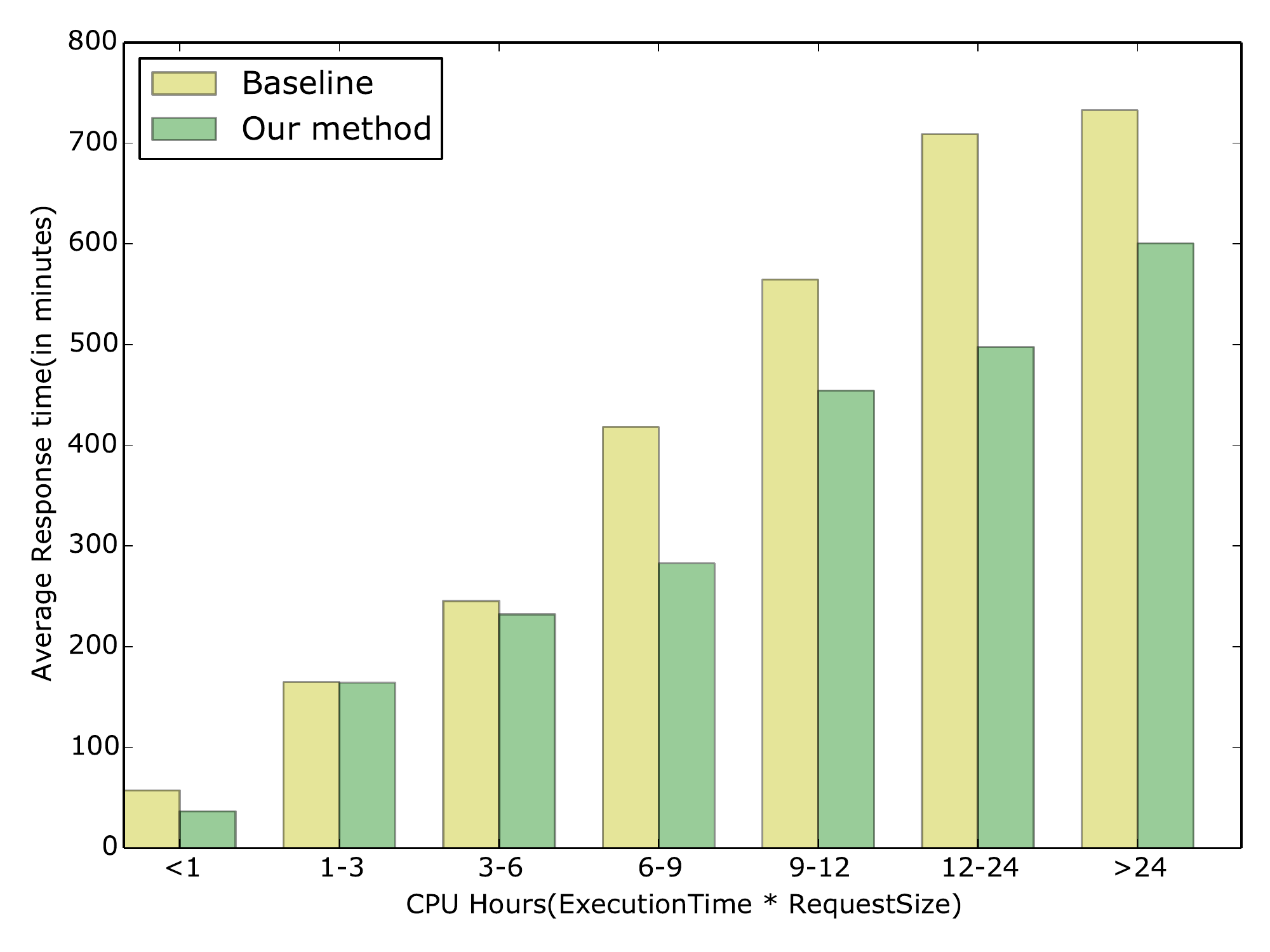}
\caption{Response Times for Jobs of Different Sizes for Delayed Submissions in SDSC SP2}
\label{cpuhours_response_ds}
\end{figure}

Figure \ref{cpuhours_response_ds} shows the average response times with the baseline and delayed submissions strategies for jobs of different sizes corresponding to different CPU hours in the SDSC SP2 trace. We find that similar to the job molding, the delayed submission strategy also uniformly provides benefits across different sizes of jobs.

\begin{figure}
\centering
  \includegraphics[scale=0.4]{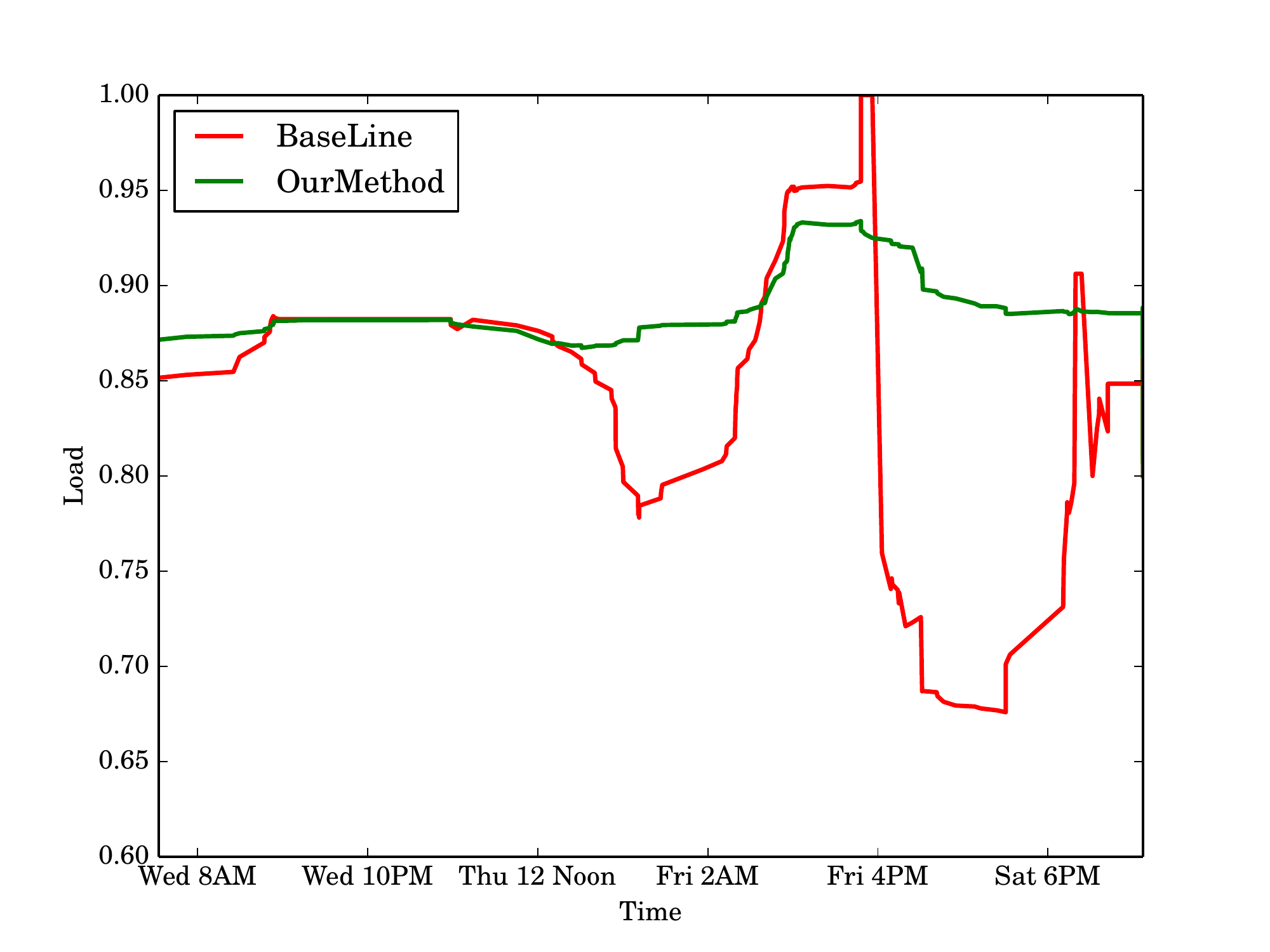}
\caption{Load Variations with Delayed Submissions in SDSC SP2}
\label{load_ds}
\end{figure}

Figure \ref{load_ds} shows the variation of load over time for the baseline and the delayed submissions strategies for a particular period of about 3.5 days in the SDSC SP2 trace. We find that the baseline method gives rise to on-peak and off-peak loads. We verified from the trace that these periods correspond to the usage during the day and night, respectively. The deep valleys exhibited by the baseline method correspond to the off-peak night hours. Our delayed submissions strategy maintains a consistent load on the system throughout the usage period. We find that our strategy moves the jobs from day executions to night executions when the system is relatively calmer. Thus, our delayed submissions strategy can automatically determine or predict these on-peak and off-peak hours to maintain a consistent load on the systems, and to obtain better response times.

\section{Related Work}
\label{related}

\textit{Predictions of Queue Waiting Times:}
QBETS \cite{nurmi-qbets-sigmetrics2007} is a forecasting system that uses quantile statistics of history submissions
to provide bounds on the queue wait times with a quantitative confidence level.  A primary limitation of QBETS is that it does not consider the state of the system, and uses only the job characteristics, which we have shown as insufficient for efficient predictions of queue wait times.

The Instance Based Learning method (IBL) by Li et al. \cite{li-qwait-ccgrid2006} considers both the job characteristics and the system states for the prediction
of queue waiting and execution times. This work uses a weighted Heterogeneous Euclidean-Overlap Distance metric to compare job attributes
and find similarities between the target job and history jobs. It then uses predictors like 1-NN (nearest neighbor), the n-WA (weighted average of n nearest neighbors), and locally weighted linear regression, and use a genetic algorithm to obtain a static weight vector which is used to improve the similarity computations.
Our queue waiting time predictions differ from this work on many aspects including
an improved representation of the system state using distributions and user-specific job submission policies, and an adaptive method to vary the prediction model for each job. We have shown that our method gives improved queue waiting time predictions over QBETS and IBL.

\textit{Predictions of Execution Times:}
There has been a large body of work on predictions of execution times using sample runs. For example, the ADEPT scalability predictor \cite{deshmeh-adept-ipdps2010} uses Downey model with a special envelope derivation technique and anomaly detection for execution time prediction using a small number of execution times (3-4) of previous executions. Most of these works assume knowledge of the particular application that is executed. Our work performs rangeSet prediction in the absence of knowledge of the particular target application executed by the user.

Smith, Foster and Taylor \cite{smith-runtimes-jpdc2004} dynamically define a set of features for defining similarity of a job with the previous jobs. They form a template set consisting of a set of features for comparison. They use genetic algorithm to search the most appropriate template set, and dynamically use a template with the smallest confidence interval for execution time prediction of a given job. The IBL method by Li et al. was also used to predict execution times and response times by considering previously submitted jobs with similar job characteristics and system states, and using weighted average methods for predictions \cite{li-ert-grid2005}.

None of these previous efforts consider the impact of load in terms of other executing jobs on the execution time of a given job. We have shown that this significantly affects execution times of potentially the same applications of same request sizes submitted by a user. Our results show that our method provides significantly better predictions of execution times over both of these efforts by Smith et al. and Li et al.

\textit{Resource Management using Predictions:}
There have been various efforts on job scheduling based on predictions using historical information. In one body of work, the predictions are used to refine the runtime estimates needed for backfilling-based schedulers. Tsafrir et al.\cite{tsafrir-backfilling-tpds2007} proposed an EASY variant that uses system-based predictions by separating the kill-time of the jobs from the runtime predictions. They use a simple predictor that uses the average runtimes of the last two jobs submitted by the same user. The PV-EASY system by Yuan et al. \cite{yuan-pveasy-hpdc2010} has developed mechanisms to use predictions and improve fairness for job scheduling. They employ a simple last model predictor and replace the user estimate with the system-generated predictions in EASY backfilling without causing reservation violations. The work by Tang et al.\cite{tang-bluegenep-ipdps2010} adjusts the user-provided ERT to improve job scheduling performance on the Blue Gene/P backfilling systems. Our work does not attempt to modify the existing scheduling algorithms and policies. Our resource management efforts treat the underlying queuing system as a black box and changes the job submission parameters, namely the request time and submission time to improve response times.

There have been some efforts that use predictions for changing the job submission parameters.
The work by Cirne and Berman \cite{cirne-herd-tpds2003} discusses job moldability for supercomputer jobs using application-level scheduler called supercomputer AppLeS (SAs). User provides the SA for his application, a set of request sizes for the job, and the SA chooses a particular request size based on application characteristics and the supercomputer current load. The load is implicitly considered in terms of the job executions of the other jobs. The work by Barsanti and Sodan \cite{barsanti-adaptivejobscheduling-jsspp2006} significantly improves the Cirne-Berman approach to also consider future job arrivals and job priorities. Similar to our work, they model load in terms of the overlap time due to other executing applications. They also seem to consider delayed submissions with a larger request size if the load at the current submission time is high. However, both of these efforts simulate the scheduling events in the system assuming a particular scheduling algorithm and queueing policy, namely, conservative or EASY backfilling, for predictions of response times. This approach inherently limits the applicability of these techniques because scheduling policies are hard to model and complete information about the scheduler is usually not published. In contrast, our proposed framework is not restricted to a specific scheduling policy and can be deployed across platforms with different resource allocation policies. Both the efforts use speedup models to predict runtime of a job for a request size that is independent of the load contributed by the other executing jobs. In our work, we have shown that different loads can result in different execution times even for the same request size. The specific impact depends on both the nature of the load and the job. Hence our work automatically derives the load functions to predict a rangeSet of runtimes.

\section{Conclusions and Future Work}
\label{con_fut}

In this work, we have developed and demonstrated an end-to-end resource management framework that uses predictions of queue waiting and execution times to minimize response times of user jobs submitted to supercomputer systems. Our resource management techniques of job molding and delayed submissions uses the predictions to reconfigure the job's request size and submission time, respectively. Using workload simulations of large supercomputer traces, we have shown large-scale improvements in predictions and reductions in response times over existing techniques and baseline strategies. We plan to explore more resource management techniques that use the predictions, and make our framework available for practical use at supercomputer centers.

\begin{small}
\bibliographystyle{IEEEtran}
\bibliography{resourceManagement}
\end{small}

\end{document}